\RequirePackage{fix-cm}
\documentclass[smallextended]{svjour3}       
\smartqed  
\usepackage{graphicx}
\usepackage{amssymb}
\usepackage[usenames,dvipsnames]{xcolor}
\usepackage[normalem]{ulem}
\begin{document}

\title{Hawking radiation as the Cosmic Censor 
}


\author{Koray D\"{u}zta\c{s}   \and     \.{I}brahim Semiz
}


\institute{K. D\"{u}zta\c{s} \and   \.{I}brahim Semiz \at
            Bo\u{g}azi\c{c}i University, Department of Physics \\ Bebek 34342, \.Istanbul, Turkey \\
              \email{koray.duztas@boun.edu.tr}           
        \and    \email{semizibr@boun.edu.tr}
}

\date{Received: date / Accepted: date}

\maketitle

\begin{abstract}
Hawking radiation acts as a cosmic censor since it carries away the angular momentum of the black hole, proportionally more than its mass.  In this work we first show that an extremal black hole cannot exist since it will be pushed away from extremality by its own Hawking radiation, without being perturbed by any external effect. We evaluate the efficiency of Hawking radiation to prevent overspinning of black holes.  We make an order of magnitude estimate to show that evaporation can prevent overspinning of black holes with  an upper limit of mass $M\lesssim 10^{17}-10^{18}\rm{g}$, when we take the interaction period to be the age of the universe. Overspinning of black holes of higher masses by test fields remains possible, even if evaporation is taken into account. We also discuss the possibility to attribute a shorter interaction period for the problem which would reduce the effect of evaporation.
\end{abstract}

\section{Introduction}
\label{intro}
Development of the singularity theorems by Penrose can be considered as the first genuine result in general relativity after Einstein. (See \cite{penrose1965rev} for a review.) The original concepts such as geodesic completeness and trapped surfaces which were introduced in this work and also used in the following extensions of singularity theorems by Penrose and Hawking \cite{singtheo}, became fundamental notions in black hole physics, cosmology, and mathematical and numerical relativity.    

Causal geodesic completeness requires that every time-like and null geodesic can be extended to arbitrarily large affine parameter value both into the future and into the past; and the most general definition of a singularity consists of its failure in a space-time region. Singularity theorems show that a space-time containing a trapped surface must possess a singularity, provided that some generic conditions are also satisfied. In classical general relativity, trapped surfaces arise in the spherically symmetric gravitational collapse of a body, thus a singularity ensues. In the model developed by Penrose and Hawking the trapped surface is contained in the black hole  region of the space-time, so it is surrounded by an event horizon.  This singularity can be considered harmless as opposed to a \emph{naked} one which intersects a Cauchy surface rendering the initial conditions undefined, thus disabling asymptotic predictability. The regions containing naked singularities also allow the evolution of closed time-like curves which violate causality \cite{tipler}. To avoid these pathologies and preserve the deterministic nature of general relativity Penrose  proposed the \emph{Cosmic Censorship Conjecture} (CCC) \cite{penrose.orig.ccc}. The weak form (WCCC) asserts that the singularities that arise in gravitational collapse are always hidden behind event horizons. Distant observers do not encounter  singularities or any effects propagating out of them. Conjecturing singularities to be inaccessible
to distant observers assures the consistency of the theory of general relativity.

It has not been possible to establish a concrete proof of CCC. Wald constructed an alternative problem to test the stability of event horizons when the black hole interacts with test particles or fields \cite{wald74}. He considered a stationary Kerr-Newman space-time uniquely defined by three parameters (Mass $M$, charge $Q$, and angular momentum per unit mass $a$), satisfying
\begin{equation}
M^{2} \geq Q^{2}+a^{2}. \label{criterion}
\end{equation}
(\ref{criterion}) is valid for  black holes surrounded by  event horizons while it is violated by naked singularities. After the black hole absorbs some particles or fields coming from infinity the space-time is expected to settle to another stationary configuration with new values of $M$, $Q$, and $a$. If it is possible to reach a final configuration of the parameters which violates (\ref{criterion}), the black hole can turn into a naked singularity and CCC is violated. Many similar attempts to check the validity of CCC can be found in literature \cite{needham,hiscock,hubeny,ri_saa_1,hod,matsasEtal,Jacobson-Sot,dkn,bck1,bck2,isoyamaEtal,ri_saa_2,toth,zimmermanEtal,gao1,gao2,kduztas,rocha,toth2}.

This problem is formulated between two stationary states in the classical context, and ignores the flux out of the black hole due to evaporation. The most general form of black hole evaporation is the Hawking radiation~\cite{hawkingrad}, which follows the discovery that certain waves are amplified during reflection from rotating black holes~\cite{zeldovich,zeldovich2}.
On a quantum particle description, amplification of waves (superradiance) corresponds to a stimulated emission of particles. However, Bogoliubov transformations formalism for particle creation does not apply to stationary space-times such as Kerr family of solutions. Therefore, Hawking considered the time dependent phase of a gravitationally collapsing body in a context that treats matter fields quantum mechanically on a classical curved space-time background. This is a good approximation to a full quantum theory of gravity outside the regions of extremely high curvature which can be encountered near a singularity. Hawking found a steady flux of particles reaching future null infinity ${\mathcal{I}}^+$. The average number of particles per unit time, with energy $\omega$ and angular momentum numbers $l,m$ is
\begin{equation}
N_{\omega l m} = \frac{\Gamma_{lm}(\omega)} { \exp [2\pi (\omega - m\Omega_{H})/\kappa]\mp 1 }
\label{hawking1}
\end{equation}
where the minus and plus signs apply to bosons and fermions respectively, $\kappa$ is the surface gravity, and $\Gamma_{lm}(\omega)$ is the fraction of a purely outgoing wave-packet of frequency $\omega$ at  ${\mathcal{I}}^+$, that would propagate through the collapsing body to  ${\mathcal{I}}^-$ when traced back in time. At sufficiently late times this equals the fraction of an ingoing wave-packet sent from  ${\mathcal{I}}^-$, that would cross the horizon of the black hole which is the analytic extension of the collapsing space-time; i.e. the fraction that would be absorbed by the black hole. Note that when a bosonic wave packet with frequency $\omega<m\Omega$ (where $\Omega=a/(r_+^2+a^2)$ is the rotational frequency of the black hole) is sent towards a Kerr black hole, the requirement that $N_{\omega l m}$ is positive implies $\Gamma_{lm}(\omega)$ is negative so that the scattered part of the wave has a larger amplitude than the original incoming wave. This is the well-known effect of superradiance exhibited by bosonic fields~\cite{zeldovich,zeldovich2}. For fermions $\Gamma_{lm}(\omega)$ remains positive for all frequencies which is in accord with the fact that they do not exhibit superradiant scattering. 

Hawking radiation as formulated in (\ref{hawking1}) refers to a higher rate of emission of particles with positive angular momentum $m$ than negative angular momentum $-m$ with the same frequency  $\omega$ and quantum number $l$. In this way the emission of particles carries away the angular momentum of the black hole \cite{hawkingrad}, working in favour of the inequality (\ref{criterion}). The essential difference between Hawking radiation and spontaneous emission by stationary black holes, first pointed out by Zel'dovich~\cite{zeldovich,zeldovich2}, then calculated by Starobinskii ~\cite{staro1,staro2} and Unruh~\cite{unruh} 
(the ``Zel'dovich-Unruh Effect'') is that emission occurs in all modes; not only in superradiant modes.  The temperature $\kappa/2\pi$ 
tends to zero for massive black holes and even for tiny ones that are nearly extremal. In this limit, Hawking radiation (\ref{hawking1}) allows the  emission of particles only in superradiant modes $\omega<m \Omega$ with $\mp \Gamma_{lm}$. This flux of particles equals to those calculated by Starobinskii  and Unruh  for a stationary Kerr black hole \cite{hawkingrad}. Superradiant modes carry higher angular momentum than energy, serving to reinforce (\ref{criterion}).  Therefore evaporation of black holes acts as a cosmic censor, considering  both  the general form of Hawking radiation and the case $\kappa \to 0$, which is more relevant for problems testing the validity of CCC.

In this work we consider a nearly extremal evaporating Kerr black hole to check if it is possible to destroy the horizon of the black hole by sending in test fields as we did in \cite{overspin}, while evaporation acts as a cosmic censor. We consider classical fields, because the energy radiated by the Hawking radiation (or the Zeldovich-Unruh effect) is many orders of magnitude larger than elementary particle masses, rendering thought experiments (\cite{ri_saa_1,matsasEtal,ri_saa_2} and others) using (quantum tunneling of) elementary particles irrelevant~\cite{CCC-SR-QPC}; and a classical field provides a good description of many particles, at least as far as energy/momentum/angular momentum flows and changes are concerned. Now the Hawking radiation, effective in this context at superradiant frequencies, act to protect WCCC, while tailored waves sent in at frequencies slightly above the superradiance limit can work towards WCCC violation. Here we make an order of magnitude estimate to evaluate the efficiency of evaporation to prevent overspinning of black holes of different masses, against the maximum effect due to challenging test fields. In the calculations we take the interaction period to be the age of the universe. We also discuss the possibility to attribute a shorter interaction period for the problem which would reduce the effect of evaporation.
\section{Evaporating Black Holes and CCC}
Consider a nearly extremal Kerr black hole at early times with parameters $M$ and $a=J/M$, satisfying
\begin{equation}
M^{2} \geq a^{2}. \label{criterion2}
\end{equation}
The black hole absorbs free fields incident from infinity where the space-time is asymptotically flat, and it keeps evaporating during the period.  
When one takes evaporation into account one has to attribute a value to the interaction period $\Delta t$ to make an estimate of the amounts of energy and angular momentum radiated away, whereas in Wald-type problems it is sufficient to assume that the interaction period is long enough. As discussed above, here we take $\Delta t$ to be the age of the universe, and also discuss the possibility to attribute a shorter interaction period for the problem. 

In fact, the steady flux of particles reaching ${\mathcal{I}}^+$ described in (\ref{hawking1}) means that the black hole will not be in a stationary state, thus can not be described by the Kerr metric. However the evaporation of the black hole is so slow that it can be described by a sequence of stationary solutions parametrized by $M$ and $a$. This quasi-stationary approximation is valid until the mass of the black hole is reduced to the Planck mass $10^{-5}$g \cite{hawkingrad}. So we formulate the problem as follows: Initially we have a nearly extremal Kerr black hole satisfying (\ref{criterion2}) and we send in massless test fields from infinity, which we define as the asymptotically flat region.  After a period $\Delta t$ the fields reach the black hole, which has been evaporating during the period. After the field interacts with the black hole we have our final configuration of $M$ and $a$. We check if the final configuration can violate (\ref{criterion2}), which is reinforced by evaporation and challenged by the incoming field.
\subsection{Challenging CCC using bosonic test fields}
In being scattered by the black hole, fields of azimuthal wave number $m$ and frequency $\omega$ cause changes in the parameters $dM$ and $dJ$, which satisfy \cite{beken}:
\begin{equation}
dJ=(m/\omega)dE \label{beken}
\end{equation} 
where $dE=dM$ for the black hole.  
The condition for CCC violation in terms of angular momentum is
\begin{equation}
\mathcal{C}_{\rm{fin}}\equiv (M+\delta E)^2-(J+\delta J)<0 \label{AngmomCond}
\end{equation} 
where $\mathcal{C}_{\rm{in}}\equiv M^2-J$. In the previous work \cite{overspin} we have used the Jacobson-Sotiriou parametrization \cite{Jacobson-Sot} for closeness to extremality in the form
\begin{equation}
J/M^2=a/M=1-2\epsilon^2, \label{eps}
\end{equation}
where $\epsilon \ll 1$ is implied; and have shown that there exists a combination of $\omega$ and $\delta E$ for any integer-spin test field incident on a slightly subextremal Kerr black hole, that can overspin the black hole into a naked singularity. The frequency has to be in the range $\omega_{\rm sl} < \omega < \omega_{1}$ where $\omega_{\rm sl}$ is the superradiance limit $m\Omega$, $\omega_1 \equiv \omega_0 /(1+\sqrt{2}\epsilon )$ and $\omega_{0} \equiv m/2M$. 
The frequency interval can be parametrized as $\omega = \omega_{0} + (s-2) \epsilon \omega_{0}$ where $0 < s < 2-\sqrt{2}$ to first order in $\epsilon.$ Then $\delta E$ must be chosen in the range delimited by the values
\begin{equation}
\delta E_{1,2} = \left[(2-s) \mp \sqrt{(2-s)^{2}-2}\right] \epsilon M,
\label{deltaE}
\end{equation}
to violate CCC. For extremal black holes ($\epsilon=0$) we have $\omega_{\rm sl}=\omega_{1}$ so the interval vanishes; therefore CCC can not be violated. The existence of a lower limit is merely due to superradiance. Had we used fermionic fields instead, the lower limit would have reduced to zero. In that case extremal black holes could also be destroyed, as long as we stay in the classical picture (see \cite{neutrinocqg} for a general discussion). However, the physical meaning of a classical fermionic field is not clear~\cite{QFT-Topo}.

The highest negative value that can be attained for $\mathcal{C}_{\rm{fin}}$ by sending in the challenging fields with $\omega_{\rm sl} < \omega < \omega_{1}$  comes from the central value of $\delta E$ in (\ref{deltaE}), that is, $\delta E_{\rm{c}}=(2-s)M\epsilon$, which gives
\begin{equation}
{\mathcal{C}_{\rm{fin}}}(\delta E=\delta E_{\rm{c}})=M^2 \epsilon^2 [2-(2-s)^2] \label{maxcont}
\end{equation}
for a given frequency in the relevant range. We may come arbitrarily close to the lower frequency limit $\omega_{\rm sl}$ ($s=0$) to get the (absolute) maximum of these values; that is $-2M^2\epsilon^2$, which is equal to $-\mathcal{C}_{\rm{in}}$! So, CCC is violated, and it is possible to even reverse the sign of $\mathcal{C}$, and not just give it some comparatively small negative value.  
\subsection{Effect of evaporation}
Because the nearly extremal black hole radiates mainly in the superradiant range, we  use the $T=\kappa/2\pi \to 0$ approximation for Hawking radiation. Scalar particles and neutrinos are produced at a similar rate \cite{unruh}, photons and gravitons are produced more copiously \cite{staro1,staro2}.The rates at which the black hole loses mass and angular momentum due to evaporation of scalar particles is given by the fluxes at infinity (see \cite{dewitt}):
\begin{eqnarray}
& &\frac{\rm{d}M}{\rm{d}t}=\lim_{r\to \infty}\int d\theta d\phi \langle T_{rt}\rangle_{\rm{vac}} \sim -\frac{e^{-\zeta}}{4\pi}\Omega^2 \label{massflux} \\
& &\frac{\rm{d}J}{\rm{d}t}=-\lim_{r\to \infty}\int d\theta d\phi \langle T_{r\phi}\rangle_{\rm{vac}} \sim -\frac{e^{-\zeta}}{2\pi}\Omega \label{angmomflux}
\end{eqnarray}
where $\zeta$ is a number of the order of unity. In \cite{CCC-SR-QPC} we hadpointed out that these fluxes make single  or few particle
gedanken experiments meaningless, and that a modified normalization
of (scalar) wave modes leads to a (semi)classical understanding of the
(scalar) Zel’dovich-Unruh effect. In that work, the question of WCCC
violation was inconclusive, since the sent-in field could not be shown to
be a small perturbation, assuming the spontaneous emission was. Here,
the sent-in field is a small perturbation as seen in (\ref{deltaE}), and we want to see the effect of evaporation.
 
For a nearly extremal black hole $J\sim M^2$, and one can make an order of magnitude estimate for the amount of angular momentum radiated away in a period $\Delta t$, assuming $M$ is almost constant and $J \sim M^2$ stays valid:
\begin{eqnarray}
& &\Delta J\sim -\frac{e^{-\zeta}}{4\pi} M^{-1}\Delta t \label{deltaj} \\
& &\Delta M\sim -\frac{e^{-\zeta}}{16\pi} M^{-2}\Delta t \label{deltam}
\end{eqnarray}
With neutrino, photon and graviton contributions included, the rate of emission is about two orders of magnitude higher than these values \cite{dewitt}. To evaluate the validity of the $J\sim M^2 \sim$ const. assumption, let us note that the expressions (\ref{massflux})-(\ref{deltam}) apply in absolute units, $G=c=\hbar=1$. In these units the mass of the Sun is $10^{38}$ and the size, the age and the mass of universe are $10^{62}$. We see that the assumption is justified for $M \gtrsim 10^{22}$, even if we take $\Delta t$ as the entire age of the universe, since then we have $\epsilon' = \vert \Delta M / M \vert \sim \vert \Delta J/ J \vert \lesssim 10^{-4}$. This lower limit corresponds to a black hole mass of $10^{17}$g, much less than stellar masses. Consider now a (nearly) extremal black hole in this mass range which has been evaporating for a certain period without being perturbed by external test particles or fields:
\begin{eqnarray}
 \mathcal{C}&=&( M + \Delta M)^2 - (J+ \Delta J) 
 = M^2 + 2M \Delta M - (J + \Delta J) \nonumber \\
&=& (M^2 - J) - \Delta J/2 
=  \mathcal{C}_{\rm_{in}} + \vert \Delta J \vert /2 \label{hawkingcc}
\end{eqnarray}
where we have used $2M \Delta M \sim \Delta J/2$ as implied by (\ref{deltaj}) and (\ref{deltam}), and neglected $(\Delta M)^{2}$. A (nearly) extremal black hole is pushed (further) away from extremality by an amount $\vert \Delta J \vert /2$ merely due to evaporation, that is, even if  an extremal black hole forms at the end of gravitational collapse, it is pushed away from extremality by its own Hawking radiation, without being perturbed by any external effect. 

Note that $\epsilon$ and $\epsilon'$ are arranged such that $\delta E \sim \epsilon M$ and $\vert \Delta M \vert = \epsilon' M$; yet $\vert \mathcal{C}_{\rm{fin}}-\mathcal{C}_{\rm{in}} \vert \sim \epsilon^{2} M^{2}$ for the incoming field, whereas $\mathcal{C}_{\rm{fin}}-\mathcal{C}_{\rm{in}} \sim \epsilon' M^{2}$ for evaporation. 
\subsection{Can evaporation prevent overspinning?}
We now consider the combination of the two processes, that is, a test field incident on an evaporating nearly extremal black hole of mass $M \gtrsim 10^{17}$g, from far away. We have
\begin{equation}
 \mathcal{C}_{\rm_{fin}}=( M + \Delta M + \delta E)^2 - (J+ \Delta J + \delta J)
\end{equation}
where $\delta E$ and $\delta J$ are perturbations due to the test field, and $\Delta M$ and $\Delta J$ denote the amount of mass and angular momentum radiated away in  the interaction period. We again start with a black hole satisfying (\ref{eps}), take $\omega = \omega_{0} / (1+a \epsilon)$, $\delta E = b \epsilon M$ and $\Delta M = - \epsilon' M$. Here, $M$ and $\epsilon$ are properties of the black hole, $a$ and $b$ are positive numbers of the order of  unity, representing the tailoring of the wave sent in, and $\epsilon'$ is determined by the black hole mass and interaction period. Both $\epsilon << 1$ and $\epsilon' << 1$ by choice of the black hole and interaction period, so the contributions to changes in $M$ and $J$ can still be calculated by the quasi-stationary approximation as above [But we do not know any a priori relation between $\epsilon$ and $\epsilon'$]. From (\ref{beken}), (\ref{deltaj}) and (\ref{deltam}) we have
\begin{equation}
\delta J = 2 (1+a \epsilon) \epsilon b M^{2}, \;\;\; \Delta J = - 4 \epsilon' M^{2} 
\end{equation}
Hence we get
\begin{equation}
\mathcal{C}_{\rm_{fin}} = 2 \epsilon' M^{2} + \epsilon'^{2} M^{2} + (b^{2}-2ab+2) \epsilon^{2} M^{2} - 2\epsilon \epsilon' b M^{2}. 
\label{eps1}
\end{equation}
i.e. the terms first order in $\epsilon$ have canceled. The second and fourth terms are negligible with respect to the first, since $\epsilon$ and $\epsilon'$ are both small. If the black hole is extremely close to extremality, we can re-parametrise its closeness as $M^2 - J = 2M^2 \epsilon''^{2}$, while $\delta E = b \epsilon M$ such that $\epsilon'' \ll \epsilon$. Then the fourth term in (\ref{eps1}) becomes $(b^{2}-2ab) \epsilon^{2} M^{2} + 2 \epsilon''^{2}M^2 \sim (b^{2}-2ab) \epsilon^{2} M^{2}$.  Now it is apparent that if $\epsilon'$ dominates $\epsilon^{2}$, $\mathcal{C}_{\rm_{fin}}$ will be positive, i.e. the overspinning is prevented; but if $\epsilon^{2}$ dominates $\epsilon'$, a $b$ can be found for every $a$ in the relevant range to make $\mathcal{C}_{\rm_{fin}}$ negative, i.e. the black hole can be overspun.

If the nearly extremal black hole that we consider has solar mass,  and even if we take $\Delta t$ as the entire age of the universe,  $\epsilon' \sim 10^{-52}$, thus evaporation has no practical effect as a cosmic censor for a black hole of solar mass against challenging test fields. If we choose $10^{-3}$ as a reasonable $\epsilon$ ($\sim 0.01\; M$ of energy in the field, one part in $10^{6}$ away from criticality!), $\epsilon'$ would have to be larger than $10^{-6}$ to prevent overspinning. With $\Delta t$ the age of the universe, we need $M < 10^{23} \sim 10^{18}g$, barely at the limit of validity of the quasi-stationary approximation. The conclusion is that for black holes more massive than $10^{17} - 10^{18}$ g, it is not very likely that evaporation can prevent violation of cosmic sensorship, if tailored fields are sent towards it. For less massive black holes, we cannot reach a conclusion, since our approximations start to break down. 

\subsection{The effect of a lower interaction period}

We have mentioned that the interaction period is the time it takes our massless fields to reach the black hole from the far away asymptotically flat region. If the object in consideration has a relatively small size one can argue that asymptotic flatness sets in at a distance much smaller than the size of the universe, and the relevant $\Delta t$ will be correpondingly smaller. This will lead to a smaller $\epsilon'$ value, decreasing the relevance of evaporation in comparison to tailored fields in our thought experiment to violate CCC. For the reasonable $\epsilon$ value discussed in the above paragraph, a decrease in $\epsilon'$ by a factor of ten would make evaporation powerless against overspinning {\it also} in the lower end ($10^{17} - 10^{18}$ g) of the mass range where our quasi-stationary approximation is valid. Hence, if the test field is sent in from a distance closer than $\sim $a billion light years, then the black hole can be over-spun despite evaporation. For example  let us consider a black hole  with $M=10^{22} \sim 10^{17}\rm{g}$ and $\epsilon = 10^{-3}$. If we send in a test field from a distance of the size of the universe, $\epsilon' \approx 10^{-4}$, so  evaporation prevents the over-spinning of this black hole. However, if the test field is sent in from $\sim 10000$ light years, then $\epsilon' \approx 10^{-10}$, and the black hole can be over-spun. 

\subsection{Using fermions or fermionic fields}
There is no superradiance for fermionic fields, so the frequency of the challenging wave does not need do be fine-tuned for CCC violation \cite{neutrinocqg}, hence is not related to $\epsilon$ any more. Therefore, for a given $\delta E$, one can increase the CCC-violating effect of the incoming field by simply decreasing the frequency, and thereby dominate the effect of evaporation.

\section{Conclusions}
In this work we have made a rough  comparison of the cosmic censorship-supporting effect of Hawking radiation  to the violating effect of challenging test fields, introduced in our previous work.

For (nearly) extremal black holes without any external perturbations, this supporting effect manifests itself as the Hawking radiation pushing the black hole away from extremality, as illustrated in (\ref{hawkingcc}). Hence even if a gravitational collapse process had resulted in an extremal black hole classically, Hawking radiation would have prevented the black hole from staying extremal.

We found that evaporation is not strong enough to prevent overspinning of black holes of mass $M\gtrsim10^{18}\rm{g}$. For smaller masses, overspinning can be prevented by evaporation if the period of interaction is sufficiently long; that is, test fields are sent into the black hole from a distance sufficiently far. Overspinning can still be achieved if test fields are sent in from a closer distance, as long as the space-time at that distance can be considered as asymptotically flat. However, for masses $M\lesssim 10^{17}\rm{g}$, our approximations break down. The overall conclusion, applicable for $M\gtrsim10^{17}\rm{g}$, is that the effect of Hawking radiation as a cosmic censor is rather weak against challenging test fields.


%
%

\begin{acknowledgements}
This work is partially supported by Bo\u{g}azi\c{c}i University Research Fund, by grant number 7981.
\end{acknowledgements}


\begin{thebibliography}{99}

\bibitem{penrose1965rev} Senovilla, J.M.M.,  Garfinkle, D.: The 1965 Penrose singularity theorem.  Class. Quantum Grav.  \textbf{32} 124008 (2015)

\bibitem{singtheo} Hawking, S.W.,  Penrose, R.: 1970 The Singularities of Gravitational Collapse and Cosmology.  Proc. R. Soc. London   \textbf{314} 529-548 (1970)

\bibitem{tipler}  Tipler,  F.J.:  Causality Violation in Asymptotically Flat Space-Times.   Phys. Rev. Lett. \textbf{37} 879-882 (1976)

\bibitem{penrose.orig.ccc} Penrose, R.:  Gravitational Collapse : The Role of General Relativity.  Riv. Nuovo Cimento \textbf{1} 252-276 (1969)

\bibitem{wald74} Wald, R.M.:  Gedanken Experiments to Destroy a Black Hole  Ann. Phys. \textbf{82} 548-556 (1974)

\bibitem{needham}Needham, T.:Cosmic Censorship and Test Particles.
       Phys. Rev. D \textbf{22} 791-796  (1980)

\bibitem{hiscock} Hiscock, W. A.:Magnetic Charge,  Black Holes and Cosmic Censorship. Ann. Phys. \textbf{131}  245-268  (1981)

\bibitem{hubeny}  Hubeny, V.E.: 1999  Overcharging a black hole and cosmic censorship. Phys. Rev. D \textbf{59} 064013 (1999)

\bibitem{ri_saa_1} Richartz, M., Saa, A.: Overspinning a Nearly Extreme Black Hole and the Weak Cosmic Censorship Conjecture. Phys. Rev. D. \textbf{78} Article No. 081503 (2008)

\bibitem{hod} Hod, S.: Return of the quantum cosmic censor. Phys. Lett. B \textbf{668} 346-349 (2008)

\bibitem{matsasEtal} Matsas, G.E.A., Richartz, M., Saa, A., da Silva A.R.R.,   Vanzella D.A.T.: Can Quantum Mechanics Fool the Cosmic Censor? Phys. Rev. D \textbf{79}  101502 (2009) 

\bibitem{Jacobson-Sot} Jacobson, T.,  Sotiriou,  T.P.:  Over-spinning a black hole with a test body. Phys. Rev. Lett. \textbf{103} 141101 (2009)

\bibitem{dkn}  Semiz, \.{I}.: 2010 Dyonic  Kerr-Newman black holes, complex scalar field and Cosmic Censorship. Gen. Relativ.  Gravit. \textbf{43} 833-846 (2010)


\bibitem{bck1} Barausse, E.,  Cardoso, V.,   Khanna, G.: Test Bodies and Naked Singularities: Is the Self-Force the Cosmic Censor?  Phys. Rev. Lett.  \textbf{105} 261102  (2010)

\bibitem{bck2} Barausse, E., Cardoso, V.,  Khanna, G:  Testing the Cosmic Censorship Conjecture with point particles: the effect of radiation reaction and the self-force. Phys. Rev. D \textbf{84} 104006 (2011) 

\bibitem{isoyamaEtal} Isoyama, S., Sago, N.,  Tanaka, T.: Cosmic Censorship in Overcharging a Reissner-Nordstr\"{o}m Black Hole via Charged Particle Absorption.  Phys. Rev. D \textbf{84}  124024 (2011)


\bibitem{ri_saa_2} Richartz, M.,   Saa,  A.:  Challenging the weak cosmic censorship conjecture with charged quantum particles   Phys. Rev. D  \textbf{84} 104021 (2011)

\bibitem{toth} Toth, G.Z.: Test of the weak cosmic censorship conjecture with a charged scalar field and dyonic Kerr-Newman black holes.  Gen. Relativ. Gravit. \textbf{44} 2019-2035 (2012) 

\bibitem{zimmermanEtal} Zimmerman, P., Vega, I.,  Poisson, E.,   Haas, R.: Self-Force as a Cosmic Censor. Phys. Rev. D  \textbf{87} 041501
 (2013)

\bibitem{gao1} Gao, S.,  Zhang, Y.:  Destroying extremal Kerr-Newman black holes with test particles. Phys. Rev. D \textbf{87} 044028 (2013)

\bibitem{gao2}  Gao, S.,  Zhang, Y.: Testing cosmic censorship conjecture near extremal black holes with cosmological constants. Int. J. Mod. Phys. D \textbf{23} 1450044 (2014)  

\bibitem{kduztas}   D\"{u}zta\c{s}, K.: Electromagnetic field and cosmic censorship. Gen. Relativ.  Gravit. \textbf{46}  1709 (2014) 

\bibitem{rocha} Rocha, J.V.:  Gravitational collapse with rotating shells and cosmic censorship. Int. J. Mod. Phys. D \textbf{24} 1542002 (2015)

\bibitem{toth2}  Toth, G.Z.: Weak cosmic censorship, dyonic Kerr-Newman black holes and Dirac fields. \textcolor{blue}{ arXiv: 1509.02878 [gr-qc] }

\bibitem{hawkingrad} Hawking, S.W.:  Particle creation by black holes. Commun. Math. Phys. \textbf{43}  199-220 (1975)

\bibitem{zeldovich} Zel'dovich, Y.B.: 1971 Generation of waves by a rotating body. JETP Lett. \textbf{14}  180-181 (1971)

\bibitem{zeldovich2}Zel'dovich, Y.B.: Amplification of cylindirical electromagnetic waves reflected from a rotating body. JETP  \textbf{35} 1085-1087  (1972)

\bibitem{staro1} Starobinski, A.A.:  Amplification of waves during reflection from a rotating black hole.  JETP \textbf{37} 28-32 (1973)

\bibitem{staro2}Starobinski, A.A.,   Churilov, S.M.:  Amplification of electromagnetic ang gravitational waves scattered by a rotating black hole   JETP \textbf{38} 1-5 (1974)

\bibitem{unruh} Unruh, W.G.:  Second quantization in the Kerr metric. Phys. Rev. D \textbf{10} 3194-3204  (1974)

\bibitem{overspin}  D\"{u}zta\c{s}, K.,   Semiz, \.{I}.:  Cosmic censorship, black holes and integer-spin test fields.  Phys. Rev. D \textbf{88} 064043 (2013)

\bibitem{CCC-SR-QPC} Semiz, \.{I}.,   D\"{u}zta\c{s}, K.: 2015 Weak Cosmic Censorship, superradiance and quantum particle creation.  Phys. Rev. D \textbf{92} 104021 (2015)

\bibitem{beken}Bekenstein, J.D.: 1973 Extraction of energy and charge from a black hole. Phys. Rev. D \textbf{7} 949-953 (1973)

\bibitem{neutrinocqg}D\"{u}zta\c{s}, K.: 2015 Stability of event horizons against neutrino
flux: the classical picture. Class. Quantum Grav. \textbf{32} 075003 (2015)

\bibitem{QFT-Topo} Schwarz, A.S.:  Quantum Field Theory and Topology, Springer, Berlin (1993)

\bibitem{dewitt} DeWitt, B.S.:  Quantum field theory in curved spacetime. Phys. Rep.  \textbf{19} 295-357 (1975)

\end{thebibliography}


%
%
\end{document}